\documentclass[aps,prd,twocolumn,amsmath,amssymb,amsfonts,nofootinbib,superscriptaddress,altaffilletter]{revtex4-1}
\pdfoutput=1

\usepackage{graphicx}
\usepackage{dcolumn}
\usepackage{bm}
\usepackage{url}
\usepackage{amssymb}
\usepackage{amsmath}
\usepackage{longtable}
\usepackage{rotating}
\usepackage{color}
\usepackage{epsfig}
\usepackage{epsf}
\usepackage{fancyhdr}
\usepackage{supertabular} 
\usepackage{hyperref}
\usepackage[raggedright,hang]{subfigure}

\newcommand{\Min}{\text{min}}
\newcommand{\ba}{\begin{align}}
\newcommand{\ea}{\end{align}}

\newcommand{\sun}{\ensuremath{\odot}}
\newcommand{\ms}{\text{M}_\odot}
\newcommand{\be}{\begin{equation}}
\newcommand{\ee}{\end{equation}}
\newcommand{\F}{\mathcal{F}}
\newcommand{\E}{\mathrm{E}}
\newcommand{\seg}{\text{seg}}
\newcommand{\Hz}{\text{Hz}}
\newcommand{\s}{\text{s}}

\newcommand{\yr}{\text{yr}}
\newcommand{\h}{\text{h}}

\newcommand{\pc}{\text{pc}}
\newcommand{\satf}{\langle2\F\rangle}
\newcommand{\msatf}{\langle2\F^\ast\rangle}

\begin{document}

\title{A directed search for Continuous Gravitational Waves from the Galactic Center}

\author{%
J.~Aasi$^{1}$,
J.~Abadie$^{1}$,
B.~P.~Abbott$^{1}$,
R.~Abbott$^{1}$,
T.~Abbott$^{2}$,
M.~R.~Abernathy$^{1}$,
T.~Accadia$^{3}$,
F.~Acernese$^{4,5}$,
C.~Adams$^{6}$,
T.~Adams$^{7}$
% ...
\\
{{}$^{\dag}$Deceased, April 2012.}
{{}$^{\ddag}$Deceased, May 2012.}
}\noaffiliation

\affiliation {LIGO - California Institute of Technology, Pasadena, CA 91125, USA }
\affiliation {Louisiana State University, Baton Rouge, LA 70803, USA }
\affiliation {Laboratoire d'Annecy-le-Vieux de Physique des Particules (LAPP), Universit\'e de Savoie, CNRS/IN2P3, F-74941 Annecy-le-Vieux, France }
\affiliation {INFN, Sezione di Napoli, Complesso Universitario di Monte S.Angelo, I-80126 Napoli, Italy }
\affiliation {Universit\`a di Salerno, Fisciano, I-84084 Salerno, Italy }
\affiliation {LIGO - Livingston Observatory, Livingston, LA 70754, USA }
\affiliation {Cardiff University, Cardiff, CF24 3AA, United Kingdom }
\affiliation {Albert-Einstein-Institut, Max-Planck-Institut f\"ur Gravitationsphysik, D-30167 Hannover, Germany }
\affiliation {Nikhef, Science Park, 1098 XG Amsterdam, The Netherlands }
\affiliation {LIGO - Massachusetts Institute of Technology, Cambridge, MA 02139, USA }
\affiliation {Instituto Nacional de Pesquisas Espaciais, 12227-010 - S\~{a}o Jos\'{e} dos Campos, SP, Brazil }
\affiliation {University of Wisconsin--Milwaukee, Milwaukee, WI 53201, USA }
\affiliation {Leibniz Universit\"at Hannover, D-30167 Hannover, Germany }
\affiliation {INFN, Sezione di Pisa, I-56127 Pisa, Italy }
\affiliation {Universit\`a di Siena, I-53100 Siena, Italy }
\affiliation {University of Florida, Gainesville, FL 32611, USA }
\affiliation {The University of Mississippi, University, MS 38677, USA }
\affiliation {California State University Fullerton, Fullerton, CA 92831, USA }
\affiliation {INFN, Sezione di Roma, I-00185 Roma, Italy }
\affiliation {University of Birmingham, Birmingham, B15 2TT, United Kingdom }
\affiliation {Albert-Einstein-Institut, Max-Planck-Institut f\"ur Gravitationsphysik, D-14476 Golm, Germany }
\affiliation {Montana State University, Bozeman, MT 59717, USA }
\affiliation {European Gravitational Observatory (EGO), I-56021 Cascina, Pisa, Italy }
\affiliation {Syracuse University, Syracuse, NY 13244, USA }
\affiliation {LIGO - Hanford Observatory, Richland, WA 99352, USA }
\affiliation {SUPA, University of Glasgow, Glasgow, G12 8QQ, United Kingdom }
\affiliation {APC, AstroParticule et Cosmologie, Universit\'e Paris Diderot, CNRS/IN2P3, CEA/Irfu, Observatoire de Paris, Sorbonne Paris Cit\'e, 10, rue Alice Domon et L\'eonie Duquet, F-75205 Paris Cedex 13, France }
\affiliation {Columbia University, New York, NY 10027, USA }
\affiliation {Stanford University, Stanford, CA 94305, USA }
\affiliation {Universit\`a di Pisa, I-56127 Pisa, Italy }
\affiliation {CAMK-PAN, 00-716 Warsaw, Poland }
\affiliation {The University of Texas at Brownsville, Brownsville, TX 78520, USA }
\affiliation {San Jose State University, San Jose, CA 95192, USA }
\affiliation {Moscow State University, Moscow, 119992, Russia }
\affiliation {LAL, Universit\'e Paris-Sud, IN2P3/CNRS, F-91898 Orsay, France }
\affiliation {NASA/Goddard Space Flight Center, Greenbelt, MD 20771, USA }
\affiliation {University of Western Australia, Crawley, WA 6009, Australia }
\affiliation {Universit\'e Nice-Sophia-Antipolis, CNRS, Observatoire de la C\^ote d'Azur, F-06304 Nice, France }
\affiliation {Institut de Physique de Rennes, CNRS, Universit\'e de Rennes 1, F-35042 Rennes, France }
\affiliation {Laboratoire des Mat\'eriaux Avanc\'es (LMA), IN2P3/CNRS, Universit\'e de Lyon, F-69622 Villeurbanne, Lyon, France }
\affiliation {Washington State University, Pullman, WA 99164, USA }
\affiliation {INFN, Sezione di Perugia, I-06123 Perugia, Italy }
\affiliation {INFN, Sezione di Firenze, I-50019 Sesto Fiorentino, Firenze, Italy }
\affiliation {Universit\`a degli Studi di Urbino 'Carlo Bo', I-61029 Urbino, Italy }
\affiliation {University of Oregon, Eugene, OR 97403, USA }
\affiliation {Laboratoire Kastler Brossel, ENS, CNRS, UPMC, Universit\'e Pierre et Marie Curie, F-75005 Paris, France }
\affiliation {Astronomical Observatory Warsaw University, 00-478 Warsaw, Poland }
\affiliation {VU University Amsterdam, 1081 HV Amsterdam, The Netherlands }
\affiliation {University of Maryland, College Park, MD 20742, USA }
\affiliation {University of Massachusetts - Amherst, Amherst, MA 01003, USA }
\affiliation {Universitat de les Illes Balears, E-07122 Palma de Mallorca, Spain }
\affiliation {Universit\`a di Napoli 'Federico II', Complesso Universitario di Monte S.Angelo, I-80126 Napoli, Italy }
\affiliation {Canadian Institute for Theoretical Astrophysics, University of Toronto, Toronto, Ontario, M5S 3H8, Canada }
\affiliation {Tsinghua University, Beijing 100084, China }
\affiliation {University of Michigan, Ann Arbor, MI 48109, USA }
\affiliation {Rochester Institute of Technology, Rochester, NY 14623, USA }
\affiliation {INFN, Sezione di Roma Tor Vergata, I-00133 Roma, Italy }
\affiliation {National Tsing Hua University, Hsinchu Taiwan 300 }
\affiliation {Charles Sturt University, Wagga Wagga, NSW 2678, Australia }
\affiliation {Caltech-CaRT, Pasadena, CA 91125, USA }
\affiliation {INFN, Sezione di Genova, I-16146 Genova, Italy }
\affiliation {Pusan National University, Busan 609-735, Korea }
\affiliation {Australian National University, Canberra, ACT 0200, Australia }
\affiliation {Carleton College, Northfield, MN 55057, USA }
\affiliation {Universit\`a di Roma Tor Vergata, I-00133 Roma, Italy }
\affiliation {Universit\`a di Roma 'La Sapienza', I-00185 Roma, Italy }
\affiliation {University of Sannio at Benevento, I-82100 Benevento, Italy and INFN (Sezione di Napoli), Italy }
\affiliation {The George Washington University, Washington, DC 20052, USA }
\affiliation {University of Cambridge, Cambridge, CB2 1TN, United Kingdom }
\affiliation {University of Minnesota, Minneapolis, MN 55455, USA }
\affiliation {The University of Sheffield, Sheffield S10 2TN, United Kingdom }
\affiliation {Wigner RCP, RMKI, H-1121 Budapest, Konkoly Thege Mikl\'os \'ut 29-33, Hungary }
\affiliation {Inter-University Centre for Astronomy and Astrophysics, Pune - 411007, India }
\affiliation {INFN, Gruppo Collegato di Trento, I-38050 Povo, Trento, Italy }
\affiliation {Universit\`a di Trento, I-38050 Povo, Trento, Italy }
\affiliation {California Institute of Technology, Pasadena, CA 91125, USA }
\affiliation {Northwestern University, Evanston, IL 60208, USA }
\affiliation {Montclair State University, Montclair, NJ 07043, USA }
\affiliation {The Pennsylvania State University, University Park, PA 16802, USA }
\affiliation {MTA-Eotvos University, \lq Lendulet\rq A. R. G., Budapest 1117, Hungary }
\affiliation {National Astronomical Observatory of Japan, Tokyo 181-8588, Japan }
\affiliation {Universit\`a di Perugia, I-06123 Perugia, Italy }
\affiliation {Rutherford Appleton Laboratory, HSIC, Chilton, Didcot, Oxon, OX11 0QX, United Kingdom }
\affiliation {Embry-Riddle Aeronautical University, Prescott, AZ 86301, USA }
\affiliation {Perimeter Institute for Theoretical Physics, Ontario, N2L 2Y5, Canada }
\affiliation {American University, Washington, DC 20016, USA }
\affiliation {University of New Hampshire, Durham, NH 03824, USA }
\affiliation {College of William and Mary, Williamsburg, VA 23187, USA }
\affiliation {University of Adelaide, Adelaide, SA 5005, Australia }
\affiliation {Raman Research Institute, Bangalore, Karnataka 560080, India }
\affiliation {Korea Institute of Science and Technology Information, Daejeon 305-806, Korea }
\affiliation {Bia{\l }ystok University, 15-424 Bia{\l }ystok, Poland }
\affiliation {University of Southampton, Southampton, SO17 1BJ, United Kingdom }
\affiliation {IISER-TVM, CET Campus, Trivandrum Kerala 695016, India }
\affiliation {Hobart and William Smith Colleges, Geneva, NY 14456, USA }
\affiliation {Institute of Applied Physics, Nizhny Novgorod, 603950, Russia }
\affiliation {Seoul National University, Seoul 151-742, Korea }
\affiliation {Hanyang University, Seoul 133-791, Korea }
\affiliation {IM-PAN, 00-956 Warsaw, Poland }
\affiliation {NCBJ, 05-400 \'Swierk-Otwock, Poland }
\affiliation {Institute for Plasma Research, Bhat, Gandhinagar 382428, India }
\affiliation {Utah State University, Logan, UT 84322, USA }
\affiliation {The University of Melbourne, Parkville, VIC 3010, Australia }
\affiliation {University of Brussels, Brussels 1050 Belgium }
\affiliation {SUPA, University of Strathclyde, Glasgow, G1 1XQ, United Kingdom }
\affiliation {ESPCI, CNRS, F-75005 Paris, France }
\affiliation {Universit\`a di Camerino, Dipartimento di Fisica, I-62032 Camerino, Italy }
\affiliation {The University of Texas at Austin, Austin, TX 78712, USA }
\affiliation {Southern University and A\&M College, Baton Rouge, LA 70813, USA }
\affiliation {IISER-Kolkata, Mohanpur, West Bengal 741252, India }
\affiliation {National Institute for Mathematical Sciences, Daejeon 305-390, Korea }
\affiliation {RRCAT, Indore MP 452013, India }
\affiliation {Tata Institute for Fundamental Research, Mumbai 400005, India }
\affiliation {Louisiana Tech University, Ruston, LA 71272, USA }
\affiliation {SUPA, University of the West of Scotland, Paisley, PA1 2BE, United Kingdom }
\affiliation {Institute of Astronomy, 65-265 Zielona G\'ora, Poland }
\affiliation {Indian Institute of Technology, Gandhinagar Ahmedabad Gujarat 382424, India }
\affiliation {Department of Astrophysics/IMAPP, Radboud University Nijmegen, P.O. Box 9010, 6500 GL Nijmegen, The Netherlands }
\affiliation {Andrews University, Berrien Springs, MI 49104, USA }
\affiliation {Trinity University, San Antonio, TX 78212, USA }
\affiliation {INFN, Sezione di Padova, I-35131 Padova, Italy }
\affiliation {University of Washington, Seattle, WA 98195, USA }
\affiliation {Southeastern Louisiana University, Hammond, LA 70402, USA }
\affiliation {Abilene Christian University, Abilene, TX 79699, USA }

\date{\today}

\begin{abstract}
We present the results of a directed search for continuous gravitational waves from unknown, isolated neutron stars in the Galactic Center region, performed on two years of data from LIGO's fifth science run from two LIGO detectors. The search uses a semi-coherent approach, analyzing coherently 630 segments, each spanning 11.5~hours, and then incoherently combining the results of the single segments. It covers gravitational wave frequencies in a range from 78 to 496~Hz and a frequency-dependent range of first order spindown values down to $-7.86\times 10^{-8}\ \Hz/\s$ at the highest frequency. No gravitational waves were detected. Placing 90\% confidence upper limits on the gravitational wave amplitude of sources at the Galactic Center, we reach $\sim 3.35\times 10^{-25}$ for frequencies near 150~Hz. These upper limits are the most constraining to date for a large-parameter-space search for continuous gravitational wave signals.
\end{abstract}

\maketitle

\section{Introduction}

In the past decade the LIGO Scientific Collaboration and the Virgo Collaboration have developed and implemented search techniques to detect gravitational wave signals. Among others, searches for continuous gravitational waves (CGWs) from known objects have been performed \cite{Abbott2010} including, for example, searches for CGWs from the low-mass X-ray binary Scorpius~X-1 \cite{Abbott2007a, Abbott2007b}, the Cas~A central compact object \cite{Wette2010} and the Crab and Vela pulsars \cite{anothercrabpaper, Abbott2008b, Abadie:2011md}. Extensive all-sky studies searching for as-yet unknown neutron stars have been performed in recent years \cite{Abbott2005a, Abbott2008a, Abbott2009a, Abbott2009c, Abbott2009d,Abadie:2011wj,Aasi:2012fw}. Because of the very weak strength of CGW signals, long integration times -- of order weeks to years -- are required to detect a signal above the noise. When the parameter space to search is large this is computationally expensive, and techniques have been developed to maximize the attainable sensitivity at fixed computing cost.

In this paper we present the first directed search for gravitational waves from yet unknown, isolated neutron stars in the direction of the Galactic Center. We use the term Galactic Center (GC) as a synonym for the coordinates of Sagittarius~A* (Sgr~A*). 
Current evolutionary scenarios predict that pulsars are born in supernova explosions of massive stars \cite{FaucherGiguere:2007zt}. 
At least three stellar clusters in the GC region contain massive stars \cite{Deneva2009} making the GC a promising target for this search. Due to the high dispersion measure toward the GC, however, out of $\sim$2000 known pulsars \cite{atnf} only six are located within 
$\sim$240~pc of Sgr~A*\cite{Johnston2006}, of which four are within 
$\sim$24 to $\sim$36~pc of Sgr~A*\cite{Deneva2009} and one magnetar is less than 2~pc away from Sgr~A*\cite{Rea:2013pqa}. 20 pulsar wind nebulae are believed to be within 20~pc from Sgr~A* \cite{Muno:2007af}. The existence of these objects supports the belief that the GC might harbor a large population of pulsars \cite{Deneva2009} not apparent to radio surveys because of the dispersion of the radio signal by galactic matter along the line of sight. 

The fact that this search targets previously unknown objects leads to a very large parameter space to be covered. A coherent search, which consists of matched filtering the data against single templates over long observation times and over a large parameter space, would have difficulty reaching an interesting sensitivity with reasonable computational power, so we resort to using a hierarchical search technique \cite{Brady1998b,Pletsch2008} which allows to integrate over the entire data set of LIGO's fifth science run (S5). 
This consists of a coherent step over shorter duration segments, using a maximum-likelihood statistic \cite{jks,Cutler:2005hc}, followed by an incoherent combination of the results from these segments.

The plan of the paper is as follows: We start with the scientific motivation of the search (Sec.~\ref{lab:motivation}) and illustrate the parameter space and the setup (Sec.~\ref{lab:paramspace}). Then we present the selection of the used data set (Sec.~\ref{lab:data}). We briefly describe the analysis method and the computational setup (Sec.~\ref{lab:anameth}). The various stages of post-processing and a coherent follow-up search are presented in Sec.~\ref{lab:postprocessing}. 
No candidate was confirmed by the follow-up. We set 90\% confidence upper limits on the GW amplitude (Sec.~\ref{lab:results}) and discuss the results in Sec.~\ref{lab:conclusion}.

\section{Motivation}
\label{lab:motivation}

Rapidly rotating neutron stars with small deviations from perfect axial symmetry are the most promising sources for continuous gravitational wave emission. No search for gravitational waves from such sources, however, has resulted in a detection yet. A possible explanation is that the detectors were not sensitive enough or that the nearest neutron stars all happen to be very close to axisymmetric. Therefore the most interesting regions are those that contain a large number of yet undiscovered neutron stars. Among such a large population it might be possible to find one neutron star that has a gravitational wave luminosity high enough or that is unusual enough to be detected with this search.\par

The GC area is believed to be such a region. The central parsec is one of the most active massive star formation regions and is believed to contain about 200 young massive stars \cite{Genzel:2010zy,Wharton:2011dv}. Because of this overabundance of massive stars, it is assumed to contain also a large number of neutron stars \cite{Johnston2006}. Massive stars are believed to be the progenitors of neutron stars: the star undergoes a supernova explosion and leaves behind the neutron star. The wide GC area ($R\le 200$~pc) 
contains more stars with initial masses above $100~\ms$ than anywhere else known in the Galaxy, plus three of the most massive young star clusters \cite{Figer:2008kf}. One of these is the central cluster, which is concentrated around the center of the Galaxy and contains at least 80 massive stars \cite{Figer:2008kf}. In the innermost $1~\pc$, the main electromagnetic radiation comes from only a few supergiants \cite{becklinneugebauer}, which are located in a dense, rich cluster, centered around Sgr~A*. Among the brightest stars we find 20 hot, massive supergiants. These stars form a sub-group concentrated strongly towards the center. The core radius of the entire central cluster is about $0.38~\pc$ \cite{unbasch1992}. 
The formation of so many massive stars in the central parsec remains a mystery \cite{Figer:2008kf}, but current estimates predict roughly as many pulsars within 0.02~pc distance to Sgr~A* as there are massive stars \cite{Pfahl:2003tf}. 
Current estimates assume at least $\sim 100$ radio pulsars to be presently orbiting Sgr~A* within this distance \cite{Pfahl:2003tf}.

\section{The search}
\subsection{The parameter space}
\label{lab:paramspace}

The targets of this search are GWs from fast spinning neutron stars with a small deviation from perfect axial symmetry. If the star rotates about its principal moment of inertia axis $I_{\text{zz}}$, the equatorial ellipticity $\epsilon$ of the neutron star is defined to be the fractional difference in the other moments of inertia,
\be
\epsilon = \frac{I_{\text{xx}}-I_{\text{yy}}}{I_{\text{zz}}}.
\ee
The amplitude of a CGW from a source emitting due to an ellipticity $\epsilon$ from a distance $r$ is \cite{jks}
\be \label{amplitude} h_0=\frac{4\pi^2 G}{c^4}\frac{I_{zz}f^2}{r}\epsilon,\ee
where $G$ is the gravitational constant, $c$ is the speed of light, and the gravitational wave frequency is twice the star's rotational frequency, $f = 2\nu$.

The range of frequencies that is covered by this search spans $78\ \Hz$ to $496\ \Hz$ and is located around the most sensitive region of the detectors (around $150\ \Hz$). Based on computational feasibility of the search, the first order spindown spans $ -f/200\ \yr \leq \dot f \leq 0\ \Hz/\s$. These ranges of frequencies and spindowns have to be covered with a set of discrete templates. The coherent analysis of the single data segments is done on a coarse rectangular grid in frequency and spindown. At the combination step the spindown parameter is refined by a factor $\gamma$ of $\mathcal{O}(1000)$. The resolutions are:
\begin{align} \label{eq:resolution}
	&\textrm{d}f = T^{-1}_\seg, \nonumber\\
	&\textrm{d}\dot f_{\text{coarse}} =  T^{-2}_\seg, \nonumber \\
	&\textrm{d}\dot f_\text{fine} = \gamma^{-1} T^{-2}_\seg, 
\end{align}
with $\gamma=3225$. This choice leads to an average mismatch\footnote{The fractional loss in detection statistic due to the finite resolution in template parameters is called mismatch.} of $\sim$0.15. In only a small fraction of cases (1\%) the mismatch could be as high as 0.4.\par

The search assumes a GW source at the position of the dynamical center of the Galaxy, the ultra compact source Sgr A* \cite{Ghez:2000ay}: 
\be \label{eq:gccoords}
\alpha = 4.650~\text{rad}~~\text{and}~~\delta = -0.506~\text{rad}.
\ee 
The angular resolution is such that the initial search is sensitive to sources within a distance $R\lesssim8~\pc$ around Sgr A*, although a coherent follow-up stage (Sec.~\ref{lab:postprocessing}) focuses on the region with $R\lesssim3~\pc$.

\subsection{The data}
\label{lab:data}

The data used for the search comes from two of the three initial LIGO ({Laser Interferometer Gravitational wave Observatory}) detectors. Initial LIGO consists of two 4-km-arm instruments in Livingston, Louisiana (L1) and Hanford, Washington State (H1) and a 2-km-long detector co-located in Hanford (H2). For this search we use data from H1 and L1 at the time of the fifth science run \cite{Abbott2009e}. The fifth science run, called S5, started on November~4th~2005 at 16:00~UTC in Hanford and on November~14th~2005 at 16:00~UTC in Livingston and ended on October~1st~2007 at 00:00~UTC. 

There exist a number of reasons for interruption of the data collection process: the detectors experience unpredictable loss of lock from seismic disturbances (earthquakes or large storms), as well as anthropogenic activities. In addition to these down-times, scheduled maintenance breaks and commissioning takes place. Some data is excluded from the analysis because of poor data quality. The remaining data is calibrated to produce a gravitational wave strain $h(t)$ time series \cite{Abbott2009d, Abbott2009e}. The time series is then broken into $1800\ \s$~long segments. Each segment is high-pass filtered above 40~Hz, Tukey windowed, and Fourier transformed to form Short Fourier Transforms (SFTs) of $h(t)$. These SFTs form the input data to our search code.\par

During S5 the detectors were operating close to or at their design sensitivity. The average strain noise of H1 and L1 was below $2.5\times10^{-23}~\text{Hz}^{-1/2}$ in the most sensitive frequency region (around 150~Hz). The performance of the detectors as well as the duty cycle improved over the course of the S5 run.\par

Our data comprises 630 segments, each spanning 11.5 hours of coincident data in H1 and L1 with the best sensitivity to a CGW signal from the GC. This setup yields the best sensitivity for given computational resources. To select the 630 segments, we use a running window of the size 11.5~h, calculate the expected SNR assuming a constant strength of the GW signal coming from the GC for the particular segment, and move the window by half an hour. To optimize sensitivity, we sort the so-obtained list of segments by their SNR values, pick the best segment, remove from the list all segments that overlap this segment and then select the next segment by taking the next on the list. This procedure is repeated until the 630th segment. 

\subsection{The analysis method}
\label{lab:anameth}

We use the hierarchical approach of \cite{Pletsch2009} (known as the global correlation transform) and divide the data into single segments, which are coherently analyzed and afterwards incoherently combined. We use the search algorithm \verb+HierarchSearchGCT+ that is part of the LAL/LALapps Software Suite \cite{lalsuite}. The coherent analysis of each single data segment is done with a matched filter technique called the $\F$-statistic \cite{jks,Cutler:2005hc}, which has been used extensively in CGW searches, most recently in \cite{Wette2010, Aasi:2012fw}. The incoherent combination step is simply a sum. What to sum, i.e. the mapping between the coarse grid and the fine grid, is described in the references provided on the global correlation transform \cite{Pletsch2008,Pletsch2009}. 

The gravitational wave amplitude $h(t)$ at the output of each detector is a linear combination of the gravitational wave functions $h_+$ and $h_\times$, where $+$ and $\times$ denote the two different polarizations of the gravitational wave signal:
\be \label{eq:respfunc} 
h(t) = F_+(t) h_+(t) + F_\times(t) h_\times (t).
\ee
$t$ is the time in the detector frame and $F_{+,\times}$ are called the antenna pattern functions \cite{jks}. $h(t)$ depends on the detector position and on the signal parameters which are: the sky location of the source of the signal, the signal's  frequency defined at the solar system barycenter at some fiducial time, its first time derivative, and four further parameters related to the amplitude and polarization: the intrinsic strain~$h_0$, the initial phase constant~$\phi_0$, the inclination angle~$\iota$ of the spin axis of the star to the line of sight, and the polarization angle~$\Psi$. These last four parameters are analytically maximized over, leaving only 
four parameters to explicitly search for: the right ascension, declination, frequency, and spindown.\par

Since the search covers only a single sky position, the right ascension and the declination are fixed to the coordinates given in Eq.~\ref{eq:gccoords}. The search templates are arranged in a rectangular grid in frequency and spindown. The result of the matched filter stage is a $2\F$ value for each segment and each template. The incoherent combination of the segments consists of summing a $2\F$ value from each segment and then dividing by the number of segments to obtain the average. By appropriately choosing which values to sum, the incoherent combination performs a refinement in spindown by a factor of $\mathcal{O}$(1000) with respect to the coherent spindown grid. The result is a value of $\satf$ for each point in this refined parameter space. The search technique does not require refinement in frequency.

The search is performed on the ATLAS cluster at the Max Planck Institute for Gravitational Physics in Hanover, Germany. The parameter space contains a total of $N = 4.4\times10^{12}$ templates and is divided among 10678 jobs, each covering a different frequency band and a range in spindown values from $ -f_\text{max}/200\ \yr \leq \dot f \leq 0\ \Hz/\s$
, where $f_\text{max}$ is the upper frequency of the band for each job. The frequency bands become smaller and smaller as the frequency increases in such a way that 
the computation time is about constant
and equal to about $\sim$5~hours on an Intel\textsuperscript{\textregistered} Xeon\textsuperscript{\textregistered} CPU X3220@2.40GHz. Each job returns the values of the detection statistic at the most significant 100,000 points in parameter space.

\subsection{Post-Processing}
\label{lab:postprocessing}

The search returns results from 1,067,800,000 points in parameter space. With the post-processing we subject these candidates to a set of vetoes aimed at removing the ones stemming from disturbances, reduce the multiplicity by clustering the ones that are not independent from one another, and zoom in on the most significant subset of these. 


The first step is the removal of all candidates that have frequencies within bands that are known to be contaminated by 
spectral artifacts. Various disturbances affect the data, like 
mechanical resonances and electrical components of the detectors, and may result in enhanced $\satf$ values. Many of these spectral disturbances are well known, 
see Tables VI and VII of \cite{{Aasi:2012fw}}, and we discard candidates that stem from the analysis of potentially contaminated data.
889,650,421 candidates survive this veto ($\sim$83.3\%).


Because of the low mismatch of the search grids, a detectable signal would produce significant values of the detection statistic in parameter space cells neighboring the actual signal location. In the second post-processing step, we cluster candidates that could be ascribed to the same signal and associate with the cluster the value of its most significant candidate. Based on results of Monte-Carlo studies we pick a fixed rectangular cluster of $2\times25$ frequency$\times$spindown bins, which is large enough to enclose parameter space cells with detection statistic values down to half of the maximum of the detection statistic of a real GW candidate.
After the application of this clustering procedure we are left with 296,815,037 candidates ($\sim$33.3\% from previous stage).


To confirm that a high $\satf$ value is the result of a GW signal, the signal must show consistent properties in the data from both detectors.
A very simple but efficient veto used in previous searches \cite{Aasi:2012fw} compares the outcome of single and multi-detector $\satf$ values  and identifies candidates stemming from local disturbances at one of the detector sites. In a false dismissal study, 500 simulated signals all passed. This veto removes $\sim$11.8\% of the candidates surviving from the previous stage.  

The next signal consistency check is computationally time-consuming and hence we do not apply it to the whole set of 261,655,549 candidates that survive up to this stage. Rather, we apply it only to the subset of candidates that could potentially show up as statistically significant in a follow-up search. 
This allows us to keep candidates whose $\satf$ value is significantly below what we expect for the loudest from the entire parameter space search on Gaussian noise\footnote{We could of course have applied this selection as a first step in the post-processing. We did not because it was the most practical to apply the signal-based vetoes described above first, and then tune the threshold for this selection based on the follow-up only.}.

%
The probability density $p^\text{loudest}(2\F^\ast \big| N)$ for the largest summed $2\F$ value over $N$ independent trials, $2\F^\ast$, is \cite{Wette2010}:
\begin{align} \label{eq:probdens}
p^\text{loudest}\big(2\F^\ast \big| N \big) &= N\ p\big(\chi^2_{4\times 630};2\F^\ast\big) \nonumber \\
&\times \left[ \int_0^{2\F^\ast}  p\big(\chi^2_{4\times 630};2\F \big) \mathrm{d}\big(2\F\big)\right]^{(N-1)},
\end{align}
where $\chi^2_{4\times 630}$ denotes a $\chi^2$-statistic with $4\times 630$ degrees of freedom. The expected value of the largest detection statistic value over $N = 4.4\times10^{12}$ independent trials simply is:
\begin{align}\label{eq:Fthr}
\E\left[2\F^\ast\right] &= \int_{0}^\infty 2\F^\ast \ p^\text{loudest}\big(2\F^\ast\big| ~4.4\times10^{12}\big) \mathrm{d}\big(2\F^\ast\big),
\end{align}
which yields a value of $4.88$ with a standard deviation of less than $0.03$.
The $N$ templates are not independent, and Eq.~\ref{eq:Fthr} slightly overestimates the actual expected value. 
A fit of the actual distribution suggests that the number of effective independent templates is $N_\text{eff} \sim {N\over 2}$. 
This moves the actual distribution of $p\left( \msatf\right)$ towards lower values of $\satf$, increasing the actual significance of candidates.
We set the threshold to $\langle2\F\rangle_\text{thr} = 4.77$ which reduces the number of candidates to 27607. The 4.77 threshold corresponds to more than 3.5 standard deviations below the expectations for the loudest over the entire search in Gaussian noise for $N_\text{eff}$ as low as ${N\over 4}$ and to 4 standard deviations below the expectations for  $N_\text{eff} \sim N$ templates .


The next veto is based on the idea that for a real signal the signal-to-noise ratio would accumulate steadily over the 630 segments, rather than be due to the high contribution of a few single segments. In contrast, noise artifacts are often limited to shorter durations in time, and hence influence the $\satf$ values only within a limited number of segments. To detect candidates with such a behavior, the average $\satf$ value is recomputed omitting the contribution from the highest $2\F$ over the 630 segments. A candidate is rejected if its recalculated $\satf$ is lower than $\langle2\F\rangle_\text{thr}$. This veto has a false dismissal rate of 0.8\% over 500 trials. 1138 candidates survive this veto. Of this set about 90\% can be ascribed to the hardware-injected pulsar 3 (see appendix \ref{lab:hardwareinjections}), leaving 59 candidates of which 20 are several standard deviations above what is expected for the loudest. Only a more sensitive follow-up search could shed light on the nature of these.


We follow-up the surviving 
59 candidates with a coherent search spanning $T_\text{seg,coh} = 90$~days of data from the H1 and L1 detectors between February~1,~2007, 15:02:57~GMT and May~2,~2007, 15:02:57~GMT. The data set was again chosen based on the sensitivity to a CGW from the GC. It contains a total of 6522 half-hour baseline SFTs (3489 from H1 and 3033 from L1), which is an average of 67.9~days from each detector. The resolution in frequency and spindown is derived from the time spanned:
\be
	\textrm{d}f_\text{coh} = (2T_{\text{seg, coh}})^{-1},~~~ \textrm{d}\dot f_\text{coh} = (2T_{\text{seg, coh}}^2)^{-1}.
\ee
The resolution in spindown turns out to be comparable to the fine grid resolution of the initial search. The frequency resolution is much finer. The covered frequency and spindown ranges are:
\begin{align}
	&\Delta f = 5\ \textrm{d}f = 5 \ T^{-1}_\seg,\nonumber\\
	&\Delta\dot f = 11\ \textrm{d}\dot f_\text{fine} = 11\ \gamma^{-1} T^{-2}_\seg,
\end{align}
centered around the frequency and spindown of the candidate to follow up. These ranges are chosen because the parameters of the highest recovered $\satf$ are always within 2 frequency bins and 5 spindown bins distance of the true signal parameters.
We are most concerned with the central 2 or 3~pc of the GC, not the entire 8~pc region covered by the initial search. 
Therefore, we concentrate on $\sim3$~pc around Sgr~A*, and place a fine template grid of 36 sky points covering a total of $7.2\times 10^{-4}$~rad in right ascension and declination, centered around it. With this setup the average mismatch of the follow-up search is 1.4\%. 
Based on the number of searched templates, the expected maximum $2\F$ for Gaussian noise is around $\sim 41\pm 3$, 
while a gravitational signal that passed the previous steps of the post-processing is expected to show up with values distributed between $\sim$50 and $\sim$400, with a mean $\sim$157 and a prominent peak at $\sim$68. Whereas it is not possible to claim a confident detection based solely on this follow-up, it is in fact possible to discard candidates as not consistent with the expectations for a signal by discarding candidates whose $2\F$ value in the coherent follow-up analysis is smaller than 50\% of the value we predict based on the $\satf$ of the original candidate. In Monte-Carlo studies with 1000 trials this procedure has a false dismissal rate of 0.4\%. The injected signal strengths were chosen such that the resulting $\satf$ values lie within the range $4.4 \lesssim \satf\lesssim 7.3$. None of the 59 candidates survives this follow-up. 

\section{Results}
\label{lab:results}

\begin{figure}
	\includegraphics[width=0.5\textwidth]{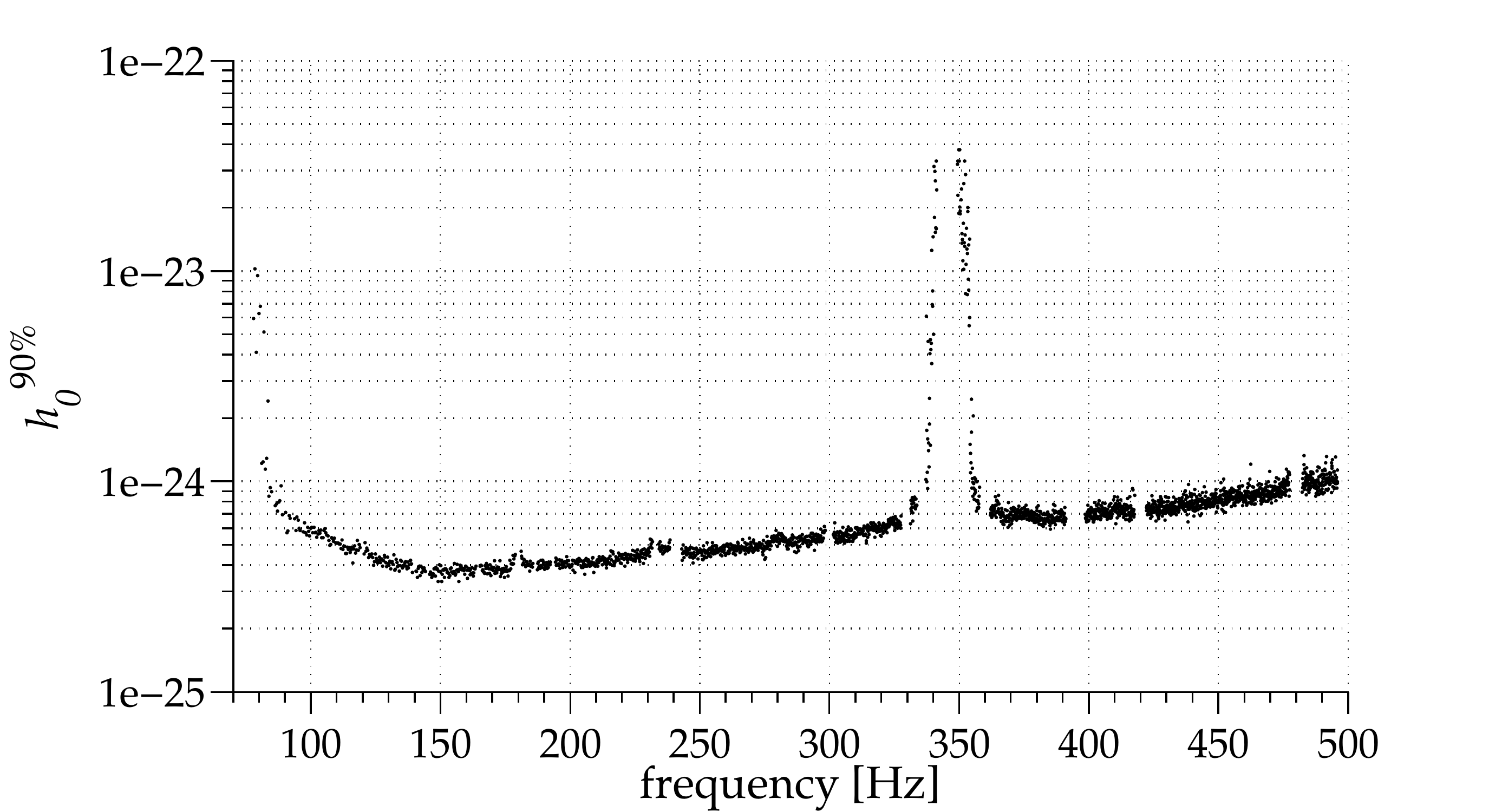}
	\caption{This plot shows the 90\% confidence upper limits on the  intrinsic GW strain $h_0$ from a population of signals with parameters within the search space. The tightest upper limit is $\sim 3.35\times 10^{-25}$ at $\sim 150$~Hz. The large value upper limit values close to 350~Hz are due to residual spectral of the detectors' violin modes.}
	\label{fig:upperlimits}
\end{figure}

We place 90\% confidence frequentist upper limits on the maximum intrinsic GW strain, $h_0^{90\%}$, from a population of signals with parameters within the search space, based on the loudest candidate from the search that could not be discarded as clearly not being of astrophysical origin. In particular, the upper limits refer to 3000 portions of the frequency-spindown parameter space with equal number of templates of about $1.5\times10^9$. They refer to sky positions within $\sim$3~pc distance of Sgr~A*, and to uniformly distributed nuisance parameters $\cos \iota$, $\phi_0$, and $\Psi$. 
$h_0^{90\%}$ is the CGW amplitude such that 90\% of a population of signals would have yielded a more significant value of the detection statistic than the most significant measured by our search in that portion of parameter space. The 90\% confidence includes the effect of different realizations of the noise in the band and of different signal shapes (different combinations of $\cos\iota, \phi_0, \Psi, f, \dot f, \alpha, \delta$) over the sub-band parameter space. This is a standard upper limit statement used in many previous searches, from \cite{Abbott:2003yq} to \cite{Aasi:2012fw}. We exclude from the upper limit statements frequency bands where more than 13\% of the parameter space was not considered due to post-processing vetoes (this reduces the UL bands to 2549 bands). The choice of 13\% was empirically determined as a good compromise between not wanting to include in the UL statements frequency bands where the searched parameter space had been significantly mutilated and wishing to keep as many valid results as possible. 
Ten further frequency bands were excluded from the upper limit statements. These bands are at neighboring frequencies to strong disturbances and themselves so disturbed that our upper limit procedure could not been applied to these frequency bands. Fig.~\ref{fig:upperlimits} shows the upper limit values. The tightest upper limit is $\sim 3.35\times 10^{-25}$ at $\sim 150$~Hz, in the spectral region where the LIGO detectors are most sensitive.

\begin{figure}
	\includegraphics[width=0.5\textwidth]{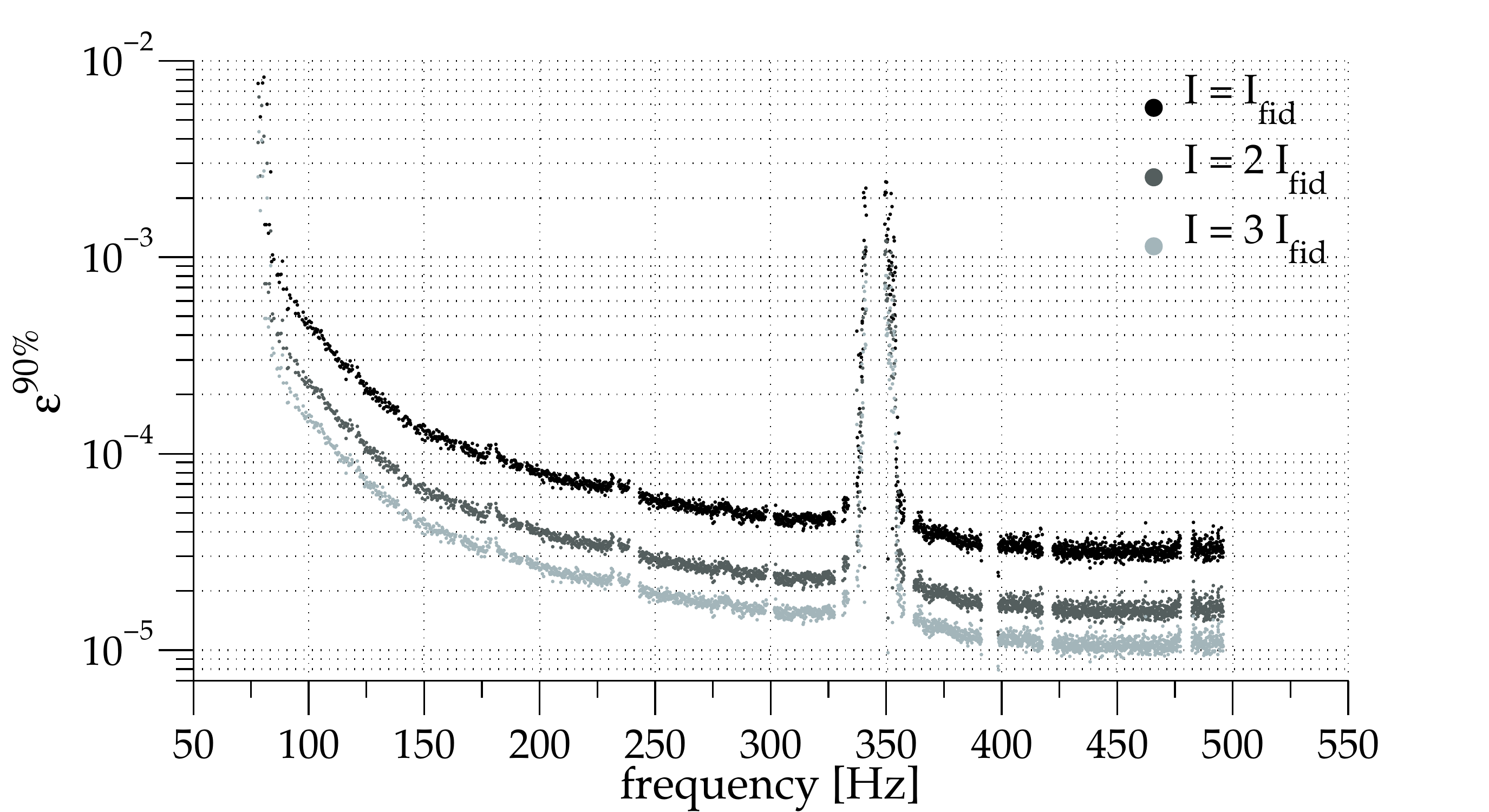}
	\caption{This plot shows the 90\% confidence upper limits on the ellipticity $\epsilon$ for our target population of sources, at a distance $r=8.3$~kpc and for three different values for the moment of inertia.} 
	\label{fig:ellipticity}
\end{figure}

Assuming a nominal value for the moment of inertia, the upper limits on $h_0$ can be recast as upper limits on the pulsar ellipticity, $\epsilon^{90\%}$. Fig.~\ref{fig:ellipticity} shows these upper limits for values of the moment of inertia between 1 and 3 times the fiducial value $I_\text{fid} = 10^{38}$~kg~m$^2$. The upper limits range from $6.2\times 10^{-3}$ at 78~Hz to $2.7\times 10^{-5}$ at 496~Hz for $I_\text{fid}$. The most constraining value is $8.7\times 10^{-6}$ at 496~Hz for $3\times I_\text{fid}$.


\begin{figure}
	\includegraphics[width=0.5\textwidth]{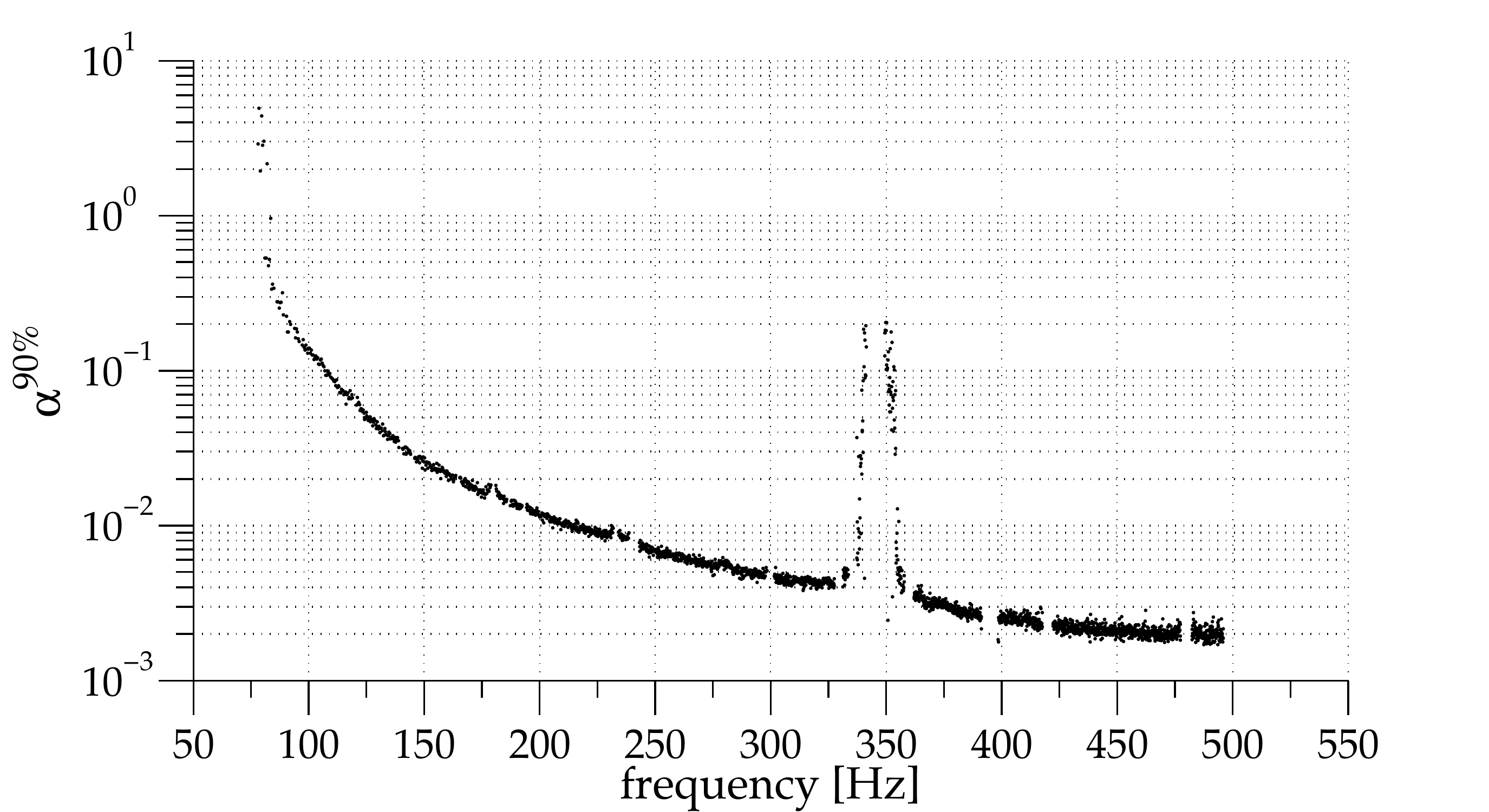}
	\caption{This plot shows the 90\% confidence upper limits on the amplitude $\alpha$ of {\it r}-mode oscillations.}
	\label{fig:rmodes}
\end{figure}

Following \cite{PhysRevD.82.104002}, the upper limits can also be translated into upper limits on the amplitude of $r$-mode oscillations, $\alpha^{90\%}$, as shown in Fig.~\ref{fig:rmodes}. The upper limits range from 2.35 at 78~Hz to 0.0016 at 496~Hz. 

\section{Conclusion}
\label{lab:conclusion}

Although this is the most sensitive directed search to date for CGWs from unknown neutron stars, no evidence for a GW signal within 3~pc of Sgr~A* was found in the searched data. 
The first upper limits on gravitational waves from the GC were set by \cite{Astone:2000jza}, a search analyzing the data of the resonant bar detector EXPLORER in the frequency range 921.32 - 921.38~Hz. The sensitivity that was reached with that search was $2.9\times 10^{-24}$. More recent upper limits on permanent signals from the GC in a wide frequency band (up to $1800\ \Hz$) were reported by \cite{radiometer}. A comparison between the results of \cite{radiometer} and the upper limits presented here is not trivial, because the upper limits set in \cite{radiometer} refer only to circular polarized waves while our results refer to an average over different polarizations. 
Also, the effect of frequency mismatch between the signal parameter and the search bins is not  folded in the results of \cite{radiometer}, whereas it is for this search. 
A further difference is that the upper limits of \cite{radiometer} are Bayesian while the results presented here are given in the Frequentist framework. Taking these differences into account, we estimate that within a 10\% uncertainty our results tighten the constraint of \cite{radiometer} by a factor of 3.2 in $h_0$.
The tightest all-sky $h_0$ upper limit in the frequency range 152.5 - 153.0~Hz from \cite{Aasi:2012fw} is $7.6\times 10^{-25}$. The results presented here tighten the \cite{Aasi:2012fw} constraint by about a factor of two. This improvement was possible because of the longer data set used, 
the higher detection efficiency of this search that targets only one point in the sky, and because of the comparatively low number of templates. For comparison, the targeted search for a CGW signal from Cas~A, which used 12 days of the same data as this search, and analyzed them with a fully coherent method, resulted in a 95\% confidence at $\sim$150~Hz of $7\times 10^{-25}$ \cite{Wette2010}. 
The improvement in sensitivity compared to the search of \cite{Wette2010} is gained by having used much more data and low-threshold post-processing. 

Following \cite{Wette:2011eu} and \cite{Aasi:2012fw} we express the GW amplitude upper limits as $h_0^{90\%} = H \sqrt{S_h/T_\text{data}}$, where $S_h$ is the detector noise and $T_\text{data} = N_\text{seg}T_\text{seg}$. The factor $H$ can be used for a direct comparison of different searches, with low values of $H$ implying, at fixed $\sqrt{S_h/T_\text{data}}$, a more effective search  \cite{analytic2f}. This search has a value of $H\sim 77$, which is an improvement of a factor 2 compared to \cite{Aasi:2012fw}, where $H$ varies within $\sim$141 and $\sim$150 with about half of the data. This confirms that the improvement in sensitivity for this search with respect to \cite{Aasi:2012fw} can be ascribed to an overall intrinsically more sensitive technique being employed, for the reasons explained above.\par

This search did not include non-zero second order spindown. This is reasonable within each coherent search segment: the largest second order spindown that over a time $T_\seg$ produces a frequency shift, $\ddot f T^2_\seg$, that is less than one half of a frequency bin is:
\be \label{eq:shift}
\ddot f T^2_{\text{seg}} \leq \frac{1}{2 T_\seg}.
\ee
Inserting $T_\seg = 11.5\ \h$, the maximum second order spindown that satisfies Eq.~\ref{eq:shift} is $\ddot f \sim 7\times 10^{-15}\ \Hz/\s^2$. Using the standard expression for the second order spindown,
\be\label{eq:spindownlimitssd}
\ddot f = n \frac{\dot f^2}{f},
\ee
and substituting $|\dot f/f| = 1/200\ \yr$, a braking index $n=5$, and $\dot f = -7.86\times 10^{-8}\ \Hz/\s$ (the largest spindown covered by the search), implies that the highest $\ddot f$ that should have been considered is $\ddot f\sim 6 \times 10^{-17}\ \Hz/\s^2$. We conclude that, for the coherent searches over $11.5$~hours, not including the second order spindown does not preclude the detection of systems in the covered search space with second order spindown values less than $\sim 6 \times 10^{-17}\ \Hz/\s^2$.
Due to the long observation time (almost two years), the second order spindown should, however, not be neglected in the incoherent combination. The minimum second order spindown signal that is necessary to move the signal by a frequency bin $\delta f$ within the observation time is:
\be\label{eq:safessd}
\ddot f_\Min = \frac{\delta f}{T_\text{obs}^2} \sim 6\times 10^{-21}\ \Hz/\s^2.
\ee
This means the presented results are surely valid for all signals with second order spindown values smaller than $6\times 10^{-21}\ \Hz/\s^2$. Computing the confidence at a fixed $h_0^{90\%}$ value for populations of signals with a second order spindown shows that signals with $\ddot f \leq 5\times 10^{-20}\ \Hz/\s^2$ do not impact the results presented in this work. This value is larger than all reliably measured values of known neutron stars as of today, where the maximum value measured is $\ddot f \simeq 1.2\times 10^{-20}\ \Hz/\s^2$ \cite{atnf}. \par
However, the standard class of signals with large spindown values is expected to also have high values of the second order spindown (Eq.~\ref{eq:spindownlimitssd}). 
Not having included a second order spindown parameter in the search means that not a standard class of objects, but rather a population with apparently very low braking indices is targeted. Such braking indices are anomalous, i.e. it would be surprising to find such objects; however they are not fundamentally impossible and could appear, for example, for stars with either a growing magnetic surface field, or a growing moment of inertia \cite{lattimer}. Under these circumstances the relationship between observed spindown and ellipticity may break down. The ellipticity of the star might be large enough that gravitational waves, even at a distance as far as the GC, can be measured at a spindown value that would not imply such strong gravitational waves in the standard picture.
This is an important fact to keep in mind when interpreting or comparing these results. 



For standard neutron stars the maximum predicted ellipticity is a few times $10^{-5}$ \cite{JohnsonMcDaniel:2012wg}. 
The upper limits on $\epsilon$ presented here are a factor of a few higher than this over most of the searched frequency band and for $I_\text{fid}$. Exotic star models do not exclude hybrid or solid stars which could sustain ellipticities up to a few $10^{-4}$ or even higher \cite{PhysRevLett.95.211101,PhysRevD.76.081502,PhysRevLett.99.231101}, well within the range that our search is sensitive to. However, since the predictions refer to the maximum values that model could sustain they don't necessarily predict those values, and hence our non-detections do not constrain the composition of neutron stars or any fundamental property of quark matter. We have considered a range of variability for the moment of inertia of the star between 1-3~$I_\text{fid}$: \cite{Thorsett:1998uc}  predicts  moments of inertia larger than $ I_\text{fid}$ for stars with masses $\ge 1 M_\sun$, which means for all neutron stars for which the masses could be measured. \cite{Bejger:2005jy} have estimated the moment of inertia for various equations of state (EOS) and predict a maximum of $I = 2.3\times I_\text{fid}$. \cite{lackey} found the highest moment of inertia to be $I = 3.3\times I_\text{fid}$ for EOS G4 in \cite{PhysRevD.73.024021}.

For frequencies in the range 50 - 500~Hz the lower limits on the distance derived in \cite{Aasi:2012fw} at the spindown limit range between 0.5 and 3.9~kpc, but because of the smaller spindown range the corresponding spindown ellipticities are lower, down to $7\times 10^{-6}$ at 500~Hz, with respect to the ellipticity upper limit values that result from this search. This reflects a different target population: closer by, and with lower ellipticities in \cite{Aasi:2012fw}; farther away, at the GC, and targeting younger stars in this analysis. We note that the $h_0^{90\%}$ upper limits presented here could also be reinterpreted as limits on different ellipticity-distance values (as done in Fig.~[13] of \cite{Abadie:2011wj}) for sources lying along the {\it direction} to the GC.

At the highest frequencies considered in this search, $\alpha^{90\%}$ reaches values which are only slightly higher than the largest ones predicted by \cite{PhysRevD.79.104003}. 
We stress that the uncertainties associated with these predictions are large enough to encompass our results.

The Advanced LIGO and Advanced Virgo detectors are expected to be operational by 2016 and to have reached their final sensitivity by 2019. The new detectors will be an order of magnitude more sensitive than the previous generation. Extrapolating from these results, a similar search on data from advanced detectors should be able to probe ellipticity values allowed for normal neutron stars at the GC and lower values for nearer objects. 

\begin{acknowledgments}
The authors gratefully acknowledge the support of the United States
National Science Foundation for the construction and operation of the
LIGO Laboratory, the Science and Technology Facilities Council of the
United Kingdom, the Max-Planck-Society, and the State of
Niedersachsen/Germany for support of the construction and operation of
the GEO600 detector, and the Italian Istituto Nazionale di Fisica
Nucleare and the French Centre National de la Recherche Scientifique
for the construction and operation of the Virgo detector. The authors
also gratefully acknowledge the support of the research by these
agencies and by the Australian Research Council, 
the International Science Linkages program of the Commonwealth of Australia,
the Council of Scientific and Industrial Research of India, 
the Istituto Nazionale di Fisica Nucleare of Italy, 
the Spanish Ministerio de Econom\'ia y Competitividad,
the Conselleria d'Economia Hisenda i Innovaci\'o of the
Govern de les Illes Balears, the Foundation for Fundamental Research
on Matter supported by the Netherlands Organisation for Scientific Research, 
the Polish Ministry of Science and Higher Education, the FOCUS
Programme of Foundation for Polish Science,
the Royal Society, the Scottish Funding Council, the
Scottish Universities Physics Alliance, The National Aeronautics and
Space Administration, 
OTKA of Hungary,
the Lyon Institute of Origins (LIO),
the National Research Foundation of Korea,
Industry Canada and the Province of Ontario through the Ministry of Economic Development and Innovation, 
the National Science and Engineering Research Council Canada,
the Carnegie Trust, the Leverhulme Trust, the
David and Lucile Packard Foundation, the Research Corporation, and
the Alfred P. Sloan Foundation. This document has been assigned LIGO Laboratory Document No. LIGO-P1300037.
\end{acknowledgments}

\appendix
\section{Hardware Injections}
\label{lab:hardwareinjections}

Over the course of the S5 run ten simulated pulsar signals were injected into the data stream by physically exciting the detectors' mirrors. Most of these fake pulsars have sky locations far away from the GC, but one of them is close enough that it contributes to the $\satf$ values of the templates in our search that are close to the injection parameters. The parameters of this hardware-injected signal are shown in Tab.~\ref{tab:pulsar3}. The distance between that hardware injection and the GC position is \mbox{$\sim1.537$~rad} in right ascension and \mbox{$\sim0.077$~rad} in declination. This is not within the covered parameter space. Nevertheless, the injected signal is so strong -- the plus- and cross-polarization translate into an implied $h_0\sim1.63\times10^{-23}$ which is a factor of $\sim 40$ louder than our $h_0^{90\%}$ at $108$ Hz -- that even a relatively marginal overlap with a template produced a significant $\satf$ value. This pulsar is detected with this search, even though it lies outside the defined parameter space.

\begin{table}[h]
	\label{tab:pulsar3}
	\begin{centering}
\begin{tabular}{rl}
\hline\hline
Value & Property\\
\hline
751680013 & Pulsar $t_\text{ref}$ in SSB frame [GPS sec]\\
$3.2766\times 10^{-20}$ & Plus-polarization signal amplitude\\
$-5.2520\times 10^{-21}$ & Cross-polarization signal amplitude\\
0.444280306 & Polarization angle psi\\
5.53 & Phase at $t_\text{ref}$ \\
108.8571594 & GW frequency at $t_\text{ref}$ [Hz]\\
-0.583578803 & Declination [rad]\\
3.113188712 & Right ascension [rad]\\
$-1.46\times10^{-17}$ & First spindown parameter [$\textrm{d}f_0/\textrm{d}t$]\\
0.0 & Second spindown parameter [$\textrm{d}f_0/\textrm{d} t^2$]\\
0.0 & Third spindown parameter [$\textrm{d}f_0/\textrm{d} t^3$]\\
\hline\hline
\end{tabular}
\caption{The parameters of the hardware injection that was detected with this search.}
\end{centering}
\end{table}

\bibliography{Bibliography}

\end{document}